\documentclass [12pt]{article}

\begin{document}
\begin{center}
\large \textbf{Entropy spectrum of a Kerr anti-de Sitter black hole}
\end{center}

\begin{center}
Deyou Chen$^{a}$\footnote{ E-mail: \underline
{ruijianchen@gmail.com} } and Haitang Yang$^{b}$\footnote{ E-mail:
\underline {hyanga@uestc.edu.cn} }
\end{center}

\begin{center}
$^{a}$Institute of Theoretical Physics, China West Normal
University, Nanchong, 637009, China

$^{b}$Department of Physics, Sichuan University, Chengdu, 610065,
China
\end{center}

\textbf{Abstract:} The entropy spectrum of a spherically symmetric
black hole was derived without the quasinormal modes in the work
of Majhi and Vagenas. Extending this work to rotating black holes,
we quantize the entropy and the horizon area of a Kerr anti-de
Sitter black hole by two methods. The spectra of entropy and area
are obtained via the Bohr-Sommerfeld quantization rule and the
adiabatic invariance in the first way. By addressing the wave
function of emitted (absorbed) particles, the entropy and the area
are quantized in the second one. Both results show that the
entropy and the area spectra are equally spaced.

\vspace*{1.0ex}
\textbf{1. Introduction}
\vspace*{1.0ex}

Quantum effects of black holes are important topics. It is widely
believed that the horizon area of black holes ought to be
quantized. In 1970s, Bekenstein first quantized the horizon area
and derived the expression of area spectrum as $A_n = n\gamma
l_p^2 $, where $\gamma $ is a dimensionless constant, $l_p $ is
the Planck length and $n = 1,2,3 \cdots $. In his work, the black
hole is non-extreme and the horizon is regarded as a classical
adiabatic invariant \cite{JDB}. Inspired by this work, people
quantized the horizon area by different methods and derived the
area spectrum which has the same form as that obtained by
Bekenstein.

One of the methods relies on quasinormal modes (QNMs), where QNMs
is defined by the solution of the perturbational wave equation
when a classical black hole is perturbed by an exterior field. The
quasinormal mode frequency plays an important role in the quantum
properties of black holes, especially in the quantization of the
horizon area. Considering this frequency denotes characteristics
of black holes, Hod employed the Bohr's correspondence principle
to quantize the horizon area and found that the area spectrum is
related to the real part of QNMs \cite{SH}. For  general
non-extreme black holes, the area spacing is $\Delta A = 4l_p^2
\ln3$. This value is in consistence with both statistical physics
arguments and the area-entropy thermodynamic relation. Latter on,
Dreyer and Kunstatter investigated the area spectra from loop
quantum gravity and the semi-classical arguments, respectively
\cite{OD,GK}. In Dreyer's work \cite{OD}, from the relation
between QNMs and the entropy spectrum, the Immirzi parameter was
fixed to $\gamma = \frac{ln3}{2\sqrt 2\pi }$. In Kunstatter's work
\cite{GK}, the vibrational frequency $\left({\omega \left( E
\right)} \right)$ of quasinormal mode is related to the energy
$\left( E \right)$ of a black hole. Then the quantity $I = \int
{\frac{dE}{\omega \left( E \right)}} $ is an adiabatic invariant
and has the equally spaced spectrum via the Bohr-Sommerfeld
quantization rule. Applying this quantity to higher-dimensional
black holes, he derived equally spaced area spectrum and entropy
spectrum of a D-dimensional Schwarzschild black hole. The area
spectra of Reissner-Nordstrom black holes, BTZ black holes and
Schwarzschild-de Sitter black holes can be referred to \cite{MRS}.

Hod's work is interesting, but there are some points needed to be
improved. First of all, the research on the asymptotic behavior of
the quasinormal mode frequencies of Kerr and Reissner-Nordstrom
black holes shows the constant $4\ln 3$ is not a universal value.
Secondly, in Hod's derivation, he only discussed transitions from
the ground state to a state with large $n$. When two arbitrary
states are taken into account, the area changes would be an
arbitrarily small value. The above problems were fixed in
\cite{MM}. It turns out that a classical black hole perturbed by
an exterior field has the same behavior as that of a damped
harmonic oscillator. Maggiore proved the physical frequency of
QNMs is determined by both its real part and imaginary part. He
further put forward that the area spectrum oughts to be determined
by the asymptotic value of the physical frequency \cite{MM}.
Applying this new explanation to the quantization of the horizon
area, he derived the equally spaced area spectrum as $A_n =
8nl_p^2 $. This value is different from that derived by Hod, but
in consistence with the result obtained by Bekenstein. Using this
new explanation, many progresses discussing the area spectra of
black holes arose \cite{ECV,AJM,KPS,BMV,ALO,SF,KR,WLLR,
KN,YSM,CYZ,SV,BCS,WLYZ,LXL,GPS}.

One method quantizing the entropy via the adiabatic invariance and
the Bohr-Sommerfeld quantization rule was put forward by Majhi and
Vagenas \cite{MV}. In their work, the adiabatic invariant quantity
connects to the Hamiltonian of black holes. Using the Hamiltonian
and the Bohr-Sommerfeld quantization rule, they derived the
equally spaced spectrum of entropy of a spherically symmetric
black hole. However, whether the horizon area spectrum is equally
spaced or not relies on the relation between the horizon area and
the entropy. The area spectrum is equally spaced when the area is
proportional to the entropy. Or, it is not equally spaced. In
\cite{ZLL}, with the observation of the periodicity of the
outgoing waves, another way was proposed to quantize the entropy.
The periodicity is exhibited by the periodic property in the wave
function. One interesting point is that QNMs is not required in
both methods.

In this paper, we adopt two ansatz to quantize the entropy and the
horizon area of a Kerr anti-de Sitter black hole. Our work in the
first ansatz is an extension of the work of Majhi and Vagenas. The
Bohr-Sommerfeld quantization rule and the adiabatic invariance are
introduced. In the second ansatz, the wave function of the emitted
(absorbed) particles plays an important role. Then the entropy and
the area are quantized by the wave function. Both results show the
entropy spectrum and the horizon spectrum are respectively equally
spaced.

The rest is organized as follows. In Sect. 2, we  derive the
Hamiltonian of the Kerr anti-de Sitter black hole with the
emission of a  particle. The quantization of the entropy and the
horizon area via the Bohr-Sommerfeld quantization rule is
dicussed. In Sect. 3, the relation between the Hamilton-Jacobi
equation and the Klein-Gordon equation is reviewed and the entropy
and the area are quantized by the wave function of particles in
the black hole's background spacetime. Sect. 4 contains
discussion and conclusion.

\vspace*{1.0ex}
\textbf{2. Quantize the entropy via the adiabatic invariance}
\vspace*{1.0ex}

In this section, the entropy and the horizon area of the Kerr anti-de Sitter
black hole are quantized via the the adiabatic invariance and the Bohr-Sommerfeld
quantization rule. The Kerr anti-de Sitter metric is given by \cite{Carter}

\begin{eqnarray}
ds^2 & = & - \frac{\Delta }{\rho ^2}\left( {dt - \frac{a}{\Xi }\sin ^2\theta
d\varphi } \right)^2 +\frac{\Delta _\theta \sin ^2\theta }{\rho ^2}\left(
{adt - \frac{r^2 + a^2}{\Xi }d\varphi } \right)^2 \nonumber\\
&& + \frac{\rho ^2}{\Delta }dr^2 + \frac{\rho ^2}{\Delta
_\theta }d\theta ^2 ,\label{eq:1}
\end{eqnarray}

\noindent
where

\begin{eqnarray*}
\Delta & = & \left( {r^2 + a^2} \right)\left( {1 + r^2l^{ - 2}} \right) - 2Mr,
\quad
\Xi = 1 - a^2l^{ - 2},
\nonumber\\
\rho ^2 & = & r^2 + a^2\cos ^2\theta , \quad \Delta _\theta = 1 -
r^2l^{ - 2}\cos ^2\theta.
\end{eqnarray*}

\noindent $a$ stands for the angular momentum per unit mass and
$l$ is a constant related to the cosmological constant $\Lambda =
- 3l^{ - 2}$. The event horizon is located at $r = r_ + $,
determined from $\Delta = 0$. The Hawking temperature and the
angular velocity at the event horizon are respectively

\begin{equation}
\Omega _ + = \frac{a\Xi }{r_ + ^2 + a^2},
\quad
T = \frac{r_ + + r_ + \left( {3r_ + ^2 + a^2} \right)l^{ - 2} - a^2r_ + ^{ -
1} }{4\pi \left( {r_ + ^2 + a^2} \right)}.\label{eq:2}
\end{equation}

The Bohr-Sommerfeld quantization rule says

\begin{equation}
\int {p_i dq_i } = nh, \label{eq:3}
\end{equation}

\noindent
where $p_i$ is the conjugate momentum of the coordinate $ q_i $ with
$ i = 0, 1, \cdots  $.
This is an adiabatic invariant quantity and $n = 1,2,3
\cdots $. To quantize the horizon area, one can first Euclideanize
the metric. The Euclideanized metric was derived by the coordinate
transformation $t \to - i\tau $ and $a \to ia $ for a general
rotating spacetime \cite{ZZM}. In this paper, we first perform the
dragging coordinate transformation, and then Euclideanize the Kerr
anti-de Sitter metric. To achieve this purpose, let $d\varphi =
-\frac{g_{03}}{g_{33}}dt$, the metric becomes

\begin{eqnarray}
ds^2  =  -\frac{\Delta\Delta_\theta \rho^2}{\Delta_\theta(r^2+a^2)^2-
\Delta a^2\sin^2}dt^2 +  \frac{\rho ^2}{\Delta }dr^2 + \frac{\rho ^2}{\Delta
_\theta }d\theta ^2 . \label{eq:4}
\end{eqnarray}

\noindent Then the Euclideanized Kerr anti-de Sitter metric is
obtained by a transformation $t \to - i\tau $. To quantize the
horizon area, we connect the adiabatic invariant quantity to the
parameters of the black hole. The adiabatic invariant quantity is
rewritten as

\begin{eqnarray}
\int {p_i dq_i } = \int {\int_0^H {\frac{d{H}'}{\dot {q}_i }dq_i } } =
\int {\int_0^H {d{H}'d\tau } } + \int {\int_0^H
{\frac{d{H}'}{\dot {x}_\mu }dx_\mu } } .\label{eq:5}
\end{eqnarray}

\noindent Now $ q_i $'s denote the coordinates of the Euclideanized
Kerr anti-de Sitter spacetime. $q_0 = \tau $ is the
Euclidean time, which has a periodicity $ T_0 = \frac{2\pi}{\kappa}$
and $\kappa $ is the surface gravity of the black hole  \cite{GWP}.
$x_\mu $'s stand for the spacelike coordinates with $\mu = 1,2$,
and $x_1 = q_1 = r $, $x_2 = q_2 = \varphi $. $H$ represents the Hamiltonian of
the system and satisfies $H = \int\limits_{\tau _i }^{\tau _f }
{Ld\tau } $, with $L$ being the Lagrangian function. Due to the
rotation of the black hole, there exists a frame dragging effect
of the coordinate system in the rotating space-time. The matter
field in the ergosphere near the horizon must be dragged by the
gravitational field with an azimuthal angular velocity. So the
Hamiltonian should reflect this effect. From the function, we find
that $\varphi $ is a cyclic coordinate and its degree of freedom
ought to be eliminated. Therefore the Hamilton function is written
as

\begin{equation}
H = \int\limits_{\tau _i }^{\tau _f } {\left( {L - P_\varphi \dot {\varphi } }
\right)d\tau } = \int\limits_{r_i }^{r_f } {\int\limits_0^{P_r } {d{P}'_r
dr} } - \int\limits_{\varphi _i }^{\varphi _f } {\int\limits_0^{P_\varphi }
{d{P}'_\varphi d\varphi } } ,\label{eq:6}
\end{equation}

\noindent where $P_r $ and $P_\varphi $ are canonical momenta of
$r$ and $\varphi $. $r_i $ and $r_f $ are locations of the event
horizons before and after the particle is emitted, respectively.
Different particles have different equations of motion. For a
massless particle, its equation of motion is the null radial
geodesics $\dot {r} = \frac{dr}{d\tau}$ \cite{PW}. While the
equation of motion is the phase velocity of the particle for a
massive one, namely $\dot {r} = v_p $ \cite{ZZ}. In this section,
we only consider the outgoing path and don't care the concrete
expression of the path in detail. Thus it does not affect our
result whether the particle is massless or not \cite{KU}. The
thermodynamic property of the Kerr anti-de Sitter black hole has
been deeply investigated. The expressions of Hamilton canonical
equations can be derived as

\begin{eqnarray}
\dot {r} & = &\left. {\frac{dH}{dP_r }} \right|_{\left( {r;\varphi ,P_\varphi }
\right)} ,
\quad
\left. {dH} \right|_{\left( {r;\varphi ,P_\varphi } \right)} = dM^\prime ;
\nonumber\\
\dot {\varphi } & = & \left. {\frac{dH}{dP_\varphi }} \right|_{\left( {\varphi ;r,P_r }
\right)} ,
\quad
\left. {dH} \right|_{\left( {\varphi ;r,P_r } \right)} = {\Omega }'
dJ^\prime ; \label{eq:7}
\end{eqnarray}

\noindent where ${\Omega }^\prime  $ and $\dot {r}$ are the
angular velocity and the equation of motion of the emitted
particle. Using Eq. (\ref{eq:7}), we can rewrite Eq. (\ref{eq:6})
as

\begin{eqnarray}
H & = & \int\limits_{r_i }^{r_f } {\int\limits_0^H {\frac{\left. {d{H}'}
\right|_{\left( {r;\varphi ,P_\varphi } \right)} }{\dot {r}}dr} } -
\int\limits_{\varphi _i }^{\varphi _f } {\int\limits_0^H {\frac{\left. {d{H}'}
\right|_{\left( {\varphi ;r,P_r ;} \right)} }{\dot {\varphi }}d\varphi } }\nonumber\\
& = & \int\limits_0^{\tau_0 } {\int\limits_0^H {\left. {d{H}'} \right|_{\left(
{r;\varphi ,P_\varphi } \right)} d\tau } } - \int\limits_0^{\tau_0 }{\int\limits_0^H
{\left. {d{H}'} \right|_{\left( {\varphi ;r,P_r } \right)} d\tau } }\nonumber\\
& = & \int\limits_0^{\tau_0} {\int\limits_0^H {d{H}'d\tau } } . \label{eq:8}
\end{eqnarray}

\noindent Now the adiabatic invariant quantity connected to the parameters of
the black hole is taken on

\begin{eqnarray}
\int {p_i dq_i } & = & 2\int\limits_0^{\tau_0}{\int_0^H {d{H}'d\tau } } \nonumber\\
& = & 2\left[ {\int\limits_0^{\tau_0} {\int\limits_0^M {dM'd\tau }
} - \int\limits_0^{\tau_0} {\int\limits_0^J {{\Omega }' d{J
'}d\tau } } } \right] .\label{eq:9}
\end{eqnarray}

\noindent Here we only consider the outgoing path. This implies we only select the
half value of periods of the Euclidean time and then $ \tau_0 = \frac{\pi}{\kappa} $.
Meanwhile from the first law of thermodynamics of the Kerr anti-de Sitter black hole

\begin{equation}
dM = TdS + \Omega _ 0 dJ,\label{eq:10}
\end{equation}

\noindent where $\Omega _ 0$ expresses the angular velocity at the
event horizon, we finish the integral and get

\begin{eqnarray}
\int {p_i dq_i }=  2\int_0^S {\tau_0 \cdot Td{S}'} = \hbar S.\label{eq:11}
\end{eqnarray}

\noindent The last equality in the above equation was obtained by
the relation between the surface gravity and the Hawking
temperature $T =\frac{\hbar \kappa}{2\pi} $. Using the
Bohr-Sommerfeld quantization rule, we get the black hole entropy
as

\begin{eqnarray}
S = 2\pi n.\label{eq:12}
\end{eqnarray}

\noindent
It shows that the entropy is discrete and the minimal spacing of the entropy
spectrum is $\Delta S = S_n - S_{n - 1} = 2\pi $, which is irrelevant to the
parameters of the Kerr anti-de Sitter  black hole. From the area-entropy
relation $S = \frac{A}{4l^2_p } $, the horizon area spectrum is obtained as

\begin{eqnarray}
A = 8\pi nl_p^2 ,\label{eq:13}
\end{eqnarray}

\noindent which implies the minimal spacing of the area spectrum
is $\Delta A = 8\pi l_p^2 $. Therefore, the entropy spectrum and
the area spectrum are respectively equally spaced and irrelevant
to the parameters of the black hole.

\vspace*{1.0ex}
\textbf{3. Quantize the entropy via the wave function }
\vspace*{1.0ex}

In this section, the entropy  and the horizon area are quantized
via the wave function of the particle in the Kerr anti-de
Sitter black hole's background spacetime. The wave function can be
obtained from the Dirac equation or the Klein-Gordon equation.
Here we choose the Klein-Gordon equation. Its connection to the
Hamilton-Jacobi equation was investigated in Ref. \cite{KM}. We
first review this connection. The Klein-Gordon equation for a
scalar field is given by

\begin{eqnarray}
g^{\mu \nu }\partial _\mu \partial _\nu \Phi - \frac{m^2}{\hbar ^2}\Phi =
0. \label{eq:14}
\end{eqnarray}

\noindent Use the WKB approximation and let the wave function have
the following form

\begin{eqnarray}
\Phi \left( {t,x^i} \right) = \exp \left[ {\frac{i}{\hbar }I\left( {t,x^i}
\right)} \right]. \label{eq:15}
\end{eqnarray}

\noindent
Inserting this expression into Eq. (\ref{eq:14}) and multiplying both sides
by $\hbar^2$ yield

\begin{eqnarray}
g^{\mu \nu }\partial _\mu I\partial _\nu I + m^2 = 0,\label{eq:16}
\end{eqnarray}

\noindent which is just the Hamilton-Jacobi equation and was used
to investigate the tunnelling radiation of scalar particles and
fermions \cite{KM,KM1}. The fermion tunnelling from the Kerr
anti-de Sitter black hole was investigated and the action $I$ of a
massive (or massless) particle was derived in \cite{DCYZ}

\begin{eqnarray}
I = - \left( {\omega - j\Omega } \right)t + W\left( r \right) + j\phi +
\Theta \left( \theta \right),\label{eq:17}
\end{eqnarray}

\noindent where $\omega = - \partial _t I$ is the energy of the
emitted particle measured by an observer at infinity and $j =
\partial _\varphi I$ is the angular quantum number about $\varphi
$. Inserting this action into Eq. (\ref{eq:15}), we get

\begin{eqnarray}
\Phi \left( {t,x^i} \right) & = & \exp \left\{ {\frac{i}{\hbar }\left[ { -
\left( {\omega - j\Omega } \right)t + W\left( r \right) + j\phi + \Theta
\left( \theta \right)} \right]} \right\}\nonumber\\
& = & \exp \left[ { - \frac{i}{\hbar }\left( {\omega - j\Omega }
\right)t} \right] \cdot \exp \left[ {\frac{i}{\hbar }W\left( r
\right)} \right] \nonumber\\
&&\cdot \exp \left( {\frac{i}{\hbar
}j\phi } \right) \cdot \exp \left[ {\frac{i}{\hbar }\Theta \left(
\theta \right)} \right]. \label{eq:18}
\end{eqnarray}

\noindent Near the horizon, $W\left( r \right) = \pm i\pi \frac{r_
+ ^2 + 2br_ + + a^2}{2\left( {r_ + - M} \right)}\left( {\omega -
j\Omega _ + } \right)$ is related to the Hawking temperature,
where the + (-) express the outgoing (ingoing) paths. In Ref.
\cite{ZLL}, the authors described the wave function in the
Schwarzschild and Kerr spacetimes and pointed out the function
$\phi$ and $\Theta(\theta)$ would vanish near the corresponding
horizons. However, from the wave function, we find they will not
vanish and have certain values, though these values will not
affect our result. Furthermore, it can be found that there are
periodicities of $t$, $\phi $ and $\Theta \left( \theta \right)$
from the expression of $\Phi \left( {t,x^i} \right)$. The periods
of them are respectively

\begin{equation}
T_0 = \frac{h}{\left( {\omega - j\Omega _ + } \right)},
\quad
\phi _0 = \frac{h}{j},
\quad
\Theta _0 \left( \theta \right) = h.\label{eq:19}
\end{equation}

\noindent The loss of a black hole system's energy is related to
the physical frequency, as addressed in Ref. \cite{MM}. In this
section, we connect the system's energy to the frequency of the
wave function and let  $E = \omega - j\Omega _+ = \hbar \omega_0
$, where $ \omega_0 $ is the frequency of the wave function. From
Eq. (\ref{eq:19}), we get the relation between the periods and the
frequency as $ T_0 = \frac{2\pi}{\omega_0}$ . Furthermore, it
should be noted that the wave function $\Phi(t,x^i)$ contains both
the outgoing and the ingoing solutions. When we consider the black
hole emits (absorbs) a particle with energy $\omega$ and angular
momentum $ j $, the changes of energy and angular momentum of this
black hole are related to these of the particle, namely $dM =
\omega $,  $dJ = j$.  From the first law of thermodynamics of this
black hole, we get

\begin{eqnarray}
dS   =  \frac{1 }{T}\left( {dM - \Omega _ + dJ} \right)
  = \frac{1 }{T}\left( {\omega - j\Omega _ + } \right)
  =  \frac{1 }{T} \frac{2 \pi \hbar }{T_0} .\label{eq:20}
\end{eqnarray}

\noindent $T$ is the temperature of the black hole, which is
related to the surface gravity as $T = \frac{\hbar \kappa}{2
\pi}$. Meanwhile, the periodic surface gravity is given by  $
\kappa = \frac{2\pi}{T_0} $ \cite{GWP}. Thus we get the change of
the black hole's entropy as

\begin{equation}
\Delta S = 2\pi ,\label{eq:21}
\end{equation}

\noindent
which shows the minimal value of change of the Kerr anti-de Sitter black
hole's entropy is $2\pi $. This result is full in consistence with that
gotten in section 2. Using the same process adopted in section 2, we can
also get the minimal spacing of the horizon area spectrum as $\Delta A = 8\pi
l_p^2 $.

\vspace*{1.0ex}
\textbf{4. Discussion and Conclusion}
\vspace*{1.0ex}

In this paper, we have quantized the entropy and the horizon area
of the Kerr anti-de Sitter black hole by two methods. In the first
one, the spectra of entropy and area were derived by the adiabatic
invariance and the Bohr-Sommerfeld quantization rule. The
adiabatic invariant quantity connects to the Hamiltonian of the
black hole. In the second ansatz, the wave function of the
particle plays an important role. The entropy was quantized by the
wave function. Both results are fully in consistence with each
other and show that the entropy spectrum and the area spectrum are
respectively equally spaced. These results are in consistence with
that obtained by Bekenstein and that derived by Maggiore. In this
paper, we used the area-entropy relation $ S = \frac {A}{4l^2_p}$,
so the area spectrum was derived and equally spaced. However, the
area is not proportional to black hole entropy in other theories,
such as Loop quantum gravity \cite{KKPM}. When we consider this
situation, the area spectrum may not be equally spaced.

\vspace*{3.0ex}
{\bf Acknowledgements}
\vspace*{1.0ex}

This work is supported in part by the Natural Science Foundation
of China (Grant No.11178018 and No. 11175039).

\end{document}